\documentclass[preprint]{aastex}
\usepackage{natbib}
\shorttitle{PAH polarization in MWC~1080 nebula}
\shortauthors{Zhang et al.}

\newcommand       \mum          {\,{\rm \mu m}}

\newcommand       \pc           {\,{\rm pc}}

\newcommand       \Lsun         {L_\sun}
\newcommand       \K            {\,{\rm K}}

\newcommand       \simali       {\sim\,}

\newcommand     \gtsim  {\lower.5ex\hbox{$\buildrel > \over \sim$}}
\newcommand     \ltsim  {\lower.5ex\hbox{$\buildrel < \over \sim$}}
\newcommand     \simgt  {\lower.5ex\hbox{$\buildrel > \over \sim$}}
\newcommand     \simlt  {\lower.5ex\hbox{$\buildrel < \over \sim$}}
 
\def 	\bE	{{\bf E}}    
\def \bJ {{\bf J}}
\def \bB {{\bf B}}

\accepted{June 5, 2017}

\begin{document}
\title{Detection of Polarized Infrared Emission 
       by Polycyclic Aromatic Hydrocarbons in the MWC~1080 Nebula        
}

\author{Han Zhang\altaffilmark{1}, Charles M. Telesco\altaffilmark{1}, Thiem Hoang\altaffilmark{2}, Aigen Li\altaffilmark{3}, Eric Pantin\altaffilmark{1,4}, Christopher M. Wright\altaffilmark{5}, Dan Li\altaffilmark{6}, Peter Barnes\altaffilmark{1,7} \\
  {\small\rm {\it Accepted by {\rm ApJ} June\ 5, 2017}}  }
 
\altaffiltext{1} {Department of Astronomy, University of Florida, Gainesville, Florida, 32611, USA, {\sl hanzh0420@ufl.edu}}
\altaffiltext{2} {Korea Astronomy and Space Science Institute 776, Daedeokdae-ro, Yuseong-gu, Daejeon 34055, Korea}
\altaffiltext{3} {Department of Physics and Astronomy, University of Missouri, Columbia, MO 65211, USA}
\altaffiltext{4} {Laboratoire AIM, CEA/DRF - CNRS - Universit\'e Paris Diderot, IRFU/DAS, 91191 Gif sur Yvette, France }
\altaffiltext{5} {School of Physical, Environmental and Mathematical Sciences, UNSW Canberra, PO Box 7916, Canberra BC 2610, Australia}
\altaffiltext{6} {Department of Physics and Astronomy, University of Pennsylvania, Philadelphia, PA 19104, USA}
\altaffiltext{7} {School of Science and Technology, University of New England, Armidale, NSW 2351, Australia}

\begin{abstract}

Polycyclic aromatic hydrocarbons (PAHs) are ubiquitous in astrophysical environments, as revealed by their pronounced
emission features at 3.3, 6.2, 7.7, 8.6, 11.3, and 12.7$\mum$ commonly ascribed to the C--H and C--C vibrational modes. Although these features have long been predicted to be polarized,
previous searches for PAH polarization led to null or, at best, tentative detections. 
Here we report the definite detection of polarized PAH emission at 11.3$\mum$
in the nebula associated with the Herbig Be star MWC 1080.
We measure a polarization degree of 1.9$\pm$0.2\%, which is unexpectedly high compared to models. This poses a challenge in the current understanding of the alignment of PAHs, which is required to polarize the PAH emission but thought to be substantially suppressed. PAH alignment with a magnetic field via a resonance paramagnetic relaxation process may account for such a high level of polarization. 

\end{abstract}


\section{Introduction}
A distinctive series of infrared (IR) emission features generally attributed to the stretching and bending vibrational modes of 
planar polycyclic aromatic hydrocarbon (PAH) 
molecules are observed in most dusty astronomical objects at 3.3, 6.2, 7.7, 8.6, 11.3, and 12.7$\mum$ \citep{1984A&A...137L...5L, 1985ApJ...290L..25A}. 
\citeauthor{1988prco.book..769L} (1988, hereafter L88) first noted that these IR emission features, if due to PAHs, are expected to be polarized as a result of anisotropic illumination by a source of ultraviolet (UV) photons (e.g., stars). 
UV absorption is favored when the molecular plane faces the illuminating source. 
If the spinning of IR-emitting PAHs do not deviate significantly from their initial orientations at UV absorption,
the IR emission will preferentially come 
from those PAHs that are facing the illuminating source 
at the time of UV absorption. 
Therefore, PAHs will emit polarized light,
with the polarization direction of out-of-plane modes (11.3, 12.7$\mum$) being along the source-molecule direction and that of in-plane modes (3.3, 6.2, 7.7, 8.6$\mum$) being perpendicular to it.
\citeauthor{2009ApJ...698.1292S} (2009, hereafter SD09) revisited the L88 scheme by considering 
realistic rotational dynamics of PAHs as well as an arbitrary degree of internal alignment between the grain symmetry axis and its angular momentum. Using realistic estimates of rotational temperatures for a typical PAH molecule of 200 carbon atoms, SD09 derived a value for the polarization fraction of $0.53\%$ for the 11.3$\mum$ feature in the case of the Orion Bar.
\cite{1988A&A...196..252S} performed the first systematic search for the polarization of the 3.3 and 11.3$\mum$ PAH features in a number of astronomical sources, obtaining upper limits of 1\% and 3\% for the 3.3 and 11.3$\mum$ features, respectively, in the Orion Bar.  

\begin{figure}
\plotone{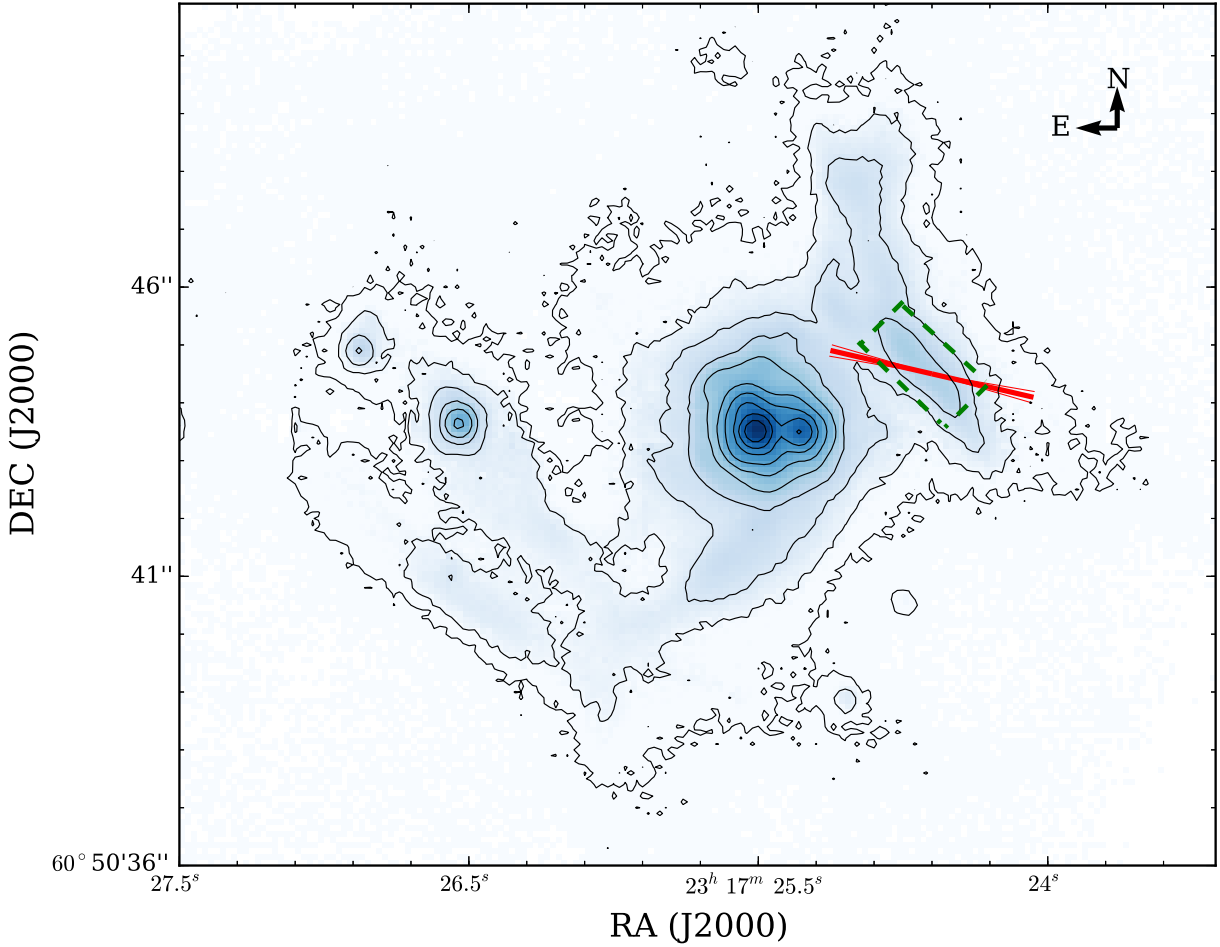}
\caption{Intensity map (contours) of MWC 1080 system at 11.2$\mum$ adopted from \cite{2014ApJ...796...74L}. The slit (dashed rectangle) is positioned to enclose the brightest part of the reflection nebula in CanariCam/GTC spectropolarimetry observations. 
 We integrate the polarization signal from the region enclosed by the dashed rectangle at 11.3$\mum$.
The thicker segment, together with the two thinner segments, shows the derived polarization position angle with 1$\sigma$ uncertainties of the 11.3$\mum$ PAH emission feature. The position angle of the projected illumination direction from the star to the nebula is $\sim$315\degr.}
\label{fig:1}
\end{figure}

To test the PAH identification of these IR features and gain insight into the alignment of PAHs, we searched for linearly polarized PAH emission in the nebula associated with MWC 1080, a stellar cluster located at a distance of 2.2 kpc 
\citep{2008ApJ...673..315W}. 
The primary star, MWC 1080A, 
is classified as a B0e star and has 
a luminosity of $L/\Lsun \approx 10^{4}$, which, together with its stellar companions, 
illuminates the surrounding gas 
and dust in the adjacent molecular cloud. 
The mid-IR image of MWC 1080 at 11.2$\mum$ resembles a pinwheel, 
with opposing arms curving off to the northwest and southeast (Fig.\,\ref{fig:1}). 
The extended mid-IR emission resembling spiral arms or wings around MWC 1080 actually traces the internal surfaces 
of a biconical cavity created by 
the outflow from MWC 1080A \citep{2014ApJ...796...74L, 2007astro.ph..1215S}. 
The brightest part of the nebula 
(green rectangle in Fig.\,\ref{fig:1}) lies 0.03$\pc$ in projection to the northwest (hereafter NW nebula) of MWC1080A
 and extends $\sim$0.1$\pc$ from the northeast to the southwest. 
That may well provide an optimal, Orion-Bar-like viewing geometry, i.e., 
a long column density along the line of sight through the photodissociation region 
and at an angle between the line of sight and illumination direction ($\alpha$$\approx$90\degr) that is almost ideal for observing maximum polarization (SD09).  

Anomalous microwave emission (AME) is often ascribed to the rotational emission from rapidly-rotating nanoparticles \citep{1998ApJ...494L..19D,1998ApJ...508..157D}. PAHs are often considered to be associated with the AME due to their abundances and small sizes \citep{1996ApJ...470..653K, 1997ApJ...486L..23L}, although there now appears to be some doubt about the hypothesis \citep{2016ApJ...827...45H}. Nitrogen-substituted PAHs (i.e., nitrogen in place of carbon), with greater dipole moments, may also be important components of the carriers of the AME \citep{2005ApJ...632..316H, 2008ApJ...680.1243M}.
Therefore, our observations have broader implications for determining the alignment and polarization of rapidly spinning PAHs, which bears on the quest for the cosmic microwave background (CMB) B-mode \citep{2013ApJ...779..152H}.

\section{Observation and Data Reduction}
CanariCam is the mid-IR (8-25$\mum$) multi-mode facility spectrometer and camera on the 10.4 m Gran Telescopio CANARIAS (GTC) in La Palma, Spain \citep{2003SPIE.4841..913T}. It employs a 320$\times$240-pixel Raytheon detector array with a pixel scale of 0$\farcs$079, which provides a field of view of 26\arcsec$\times$19\arcsec with Nyquist sampling (two pixels per $\lambda$/D) of the diffraction-limited point-spread function at 8$\mum$.
Polarimetry is accomplished through the use of a Wollaston prism and a half-wave plate rotated to angles of 0\degr, 22.5\degr, 45\degr, and 67.5\degr. A Wollaston prism in the optical path divides incoming light into two beams (ordinary and extraordinary), which are recorded by the detector simultaneously.

We obtained low-resolution (R$\equiv$$\lambda$/$\Delta\lambda$$\approx$50) spectropolarimetry observations of the NW nebula of MWC 1080 on 2015 July 31, August 5, and August 7 spanning the wavelength range 7.5--13.0$\mum$. We made four separate measurements, as indicated in the observation log presented in Table \ref{tab:1}. 
The spectroscopic observations of the NW nebula were interlaced with observations of the Cohen standard star HD 21330 \citep{1999AJ....117.1864C} for flux and point-spread-function (PSF) calibration, and the standard star AFGL 2591, selected from \cite{2000MNRAS.312..327S} to calibrate the polarization position angle. 
The standard mid-IR chop-nod technique was applied with an 11$\arcsec$ chop throw in the northwest-southeast direction. 
We positioned the 1$\farcs$04$\times$2$\farcs$08 slit 
with the slit's longer axis oriented at 45$\degr$ from the North to intersect the brightest part of the nebula, enclosed by the green rectangle in Fig.\,\ref{fig:1}. The 11.2$\mum$ image of MWC 1080 is adopted from \cite{2014ApJ...796...74L}, and the data were taken using Michelle, the facility mid-IR camera at Gemini North.

The data were reduced using custom IDL software, as described in \cite{2014ascl.soft05014B} and \cite{2014ascl.soft11009L}. 
We extracted one-dimensional spectra by integrating the central 21 pixels (1$\farcs$6) along the slit direction. 
Wavelength calibration was done using several sky lines identified in the raw images. We computed normalized Stokes parameters \textit{q} (\textit{q=Q/I}) and \textit{u} (\textit{u=U/I}) using both the difference and ratio methods, with each providing consistent results \citep{2005aspo.book.....T}.
The data were calibrated for instrumental efficiency and polarization.
The estimated instrumental polarization was 0.6\% as measured with HD 21330 and was subtracted from the observations of the NW nebula in the $Q-U$ plane. 
The degree of polarization $p=\sqrt{q^{2}+u^{2}-\sigma^{2}}$, where the last term (the `debias' term) was introduced to remove a positive offset in the signal floor resulting from squared background noise. Debiasing may introduce negative values if the noise fluctuations are stronger than the signal. The polarization position angle was computed as $\theta=0.5{\rm arctan}(u/q)$. 
The uncertainties $\sigma_{q}$ and $\sigma_{u}$ associated with the normalized Stokes parameters were derived using a standard 3-sigma clipping algorithm \citep{1987igbo.conf.....R}, and were then propagated through the analysis to obtain the polarization uncertainty $\sigma_{p}$ and position angle uncertainty $\sigma_{\theta}=\sigma_{p}/2p$ \citep{2006PASP..118..146P}. All the polarization position angles were calibrated east from north.
We masked out the region between 9.2--10.0$\mum$, which is dominated by the atmospheric ozone feature.
The three-pixel (0.06$\mum$) boxcar-smoothed intensity spectrum of the nebula (Stokes I) is presented in Fig.\,\ref{fig:2}a. 

To further increase the signal-to-noise ratio (S/N) of the polarization measurements, we rebinned the ordinary and extraordinary ray spectra into 0.12$\mum$ wavelength (6 pixels) bins (downsampling) and then applied an additional three-pixel (0.36$\mum$) boxcar-smoothing to the data. That results in an equivalent spectral resolution of the polarization spectrum of R$\approx$32 (Fig.\,\ref{fig:2}b).  
Stokes $u$ and $q$ are plotted in Figs.\,\ref{fig:3}b and c, respectively.  
To emphasize the statistical significance of the measurements, we plot the S/N of the polarized intensity in Fig.\,\ref{fig:3}a.

\section{Results}

\begin{figure}
\plotone{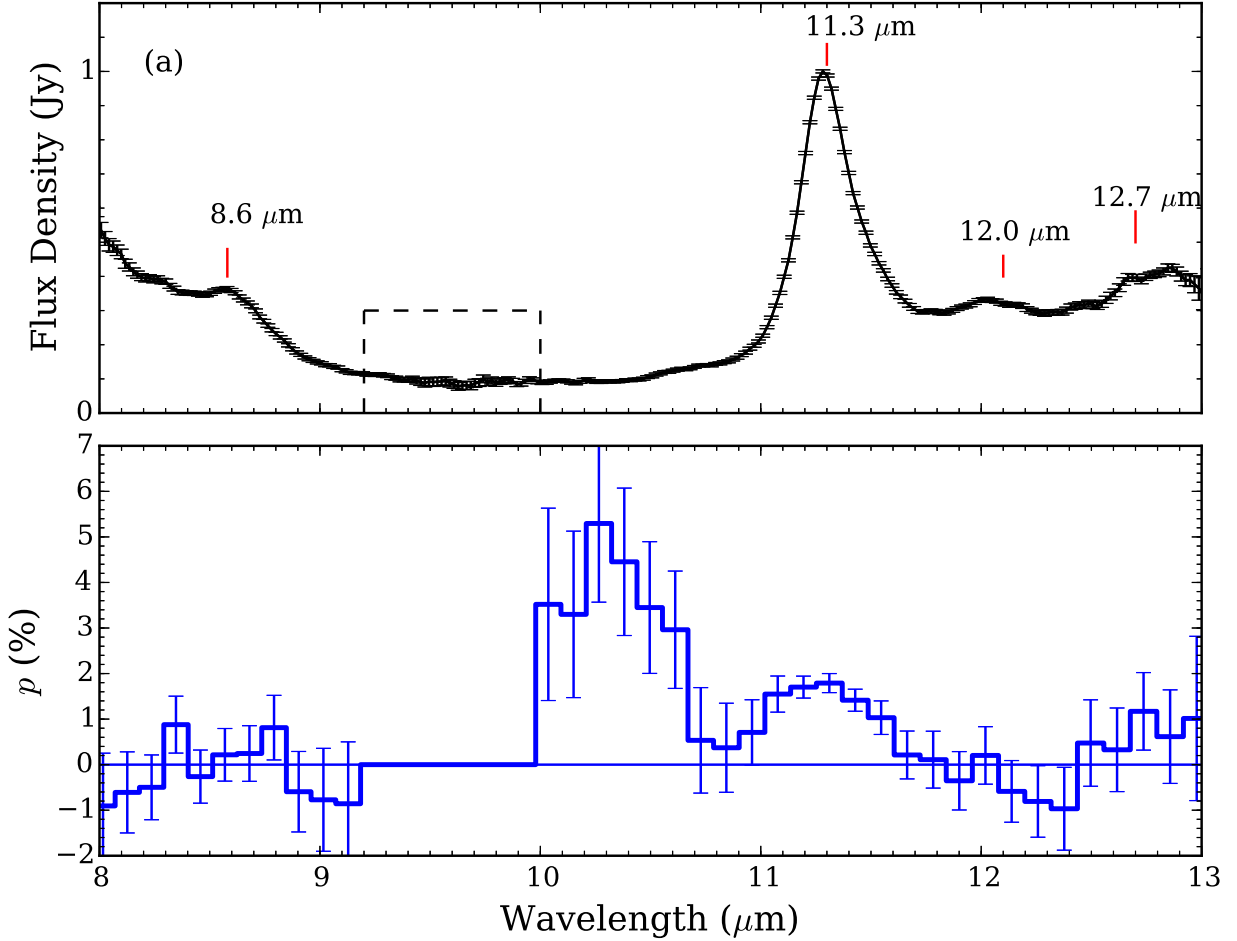}
\caption{(a): Canaricam/GTC low-resolution (R$\approx$50) spectrum of the brightest central 1$\farcs$6 (21 pixels) region of the NW nebula. 
The raw data were smoothed with a boxcar of width 3 pixels (0.06$\mum$). PAH emission features are seen at 8.6, 11.3, 12.0, and 12.7$\mum$, in which the 8.6$\mum$ feature originates from C-H in-plane bending modes and the 11.3, 12.0, 12.7$\mum$ features originate from C-H out-of-plane bending modes.
(b): Polarization percentage \textit{p} of the NW nebula. The polarization data were downsampled by 6 pixels (0.12$\mum$) and then smoothed with a boxcar of width 3 pixels (0.36$\mum$). The final spectral resolution R$\approx$32. We masked out the region between 9.2--10.0$\mum$ which is dominated by the atmospheric ozone feature (the dashed box).}
\label{fig:2}
\end{figure}

\begin{figure}
\plotone{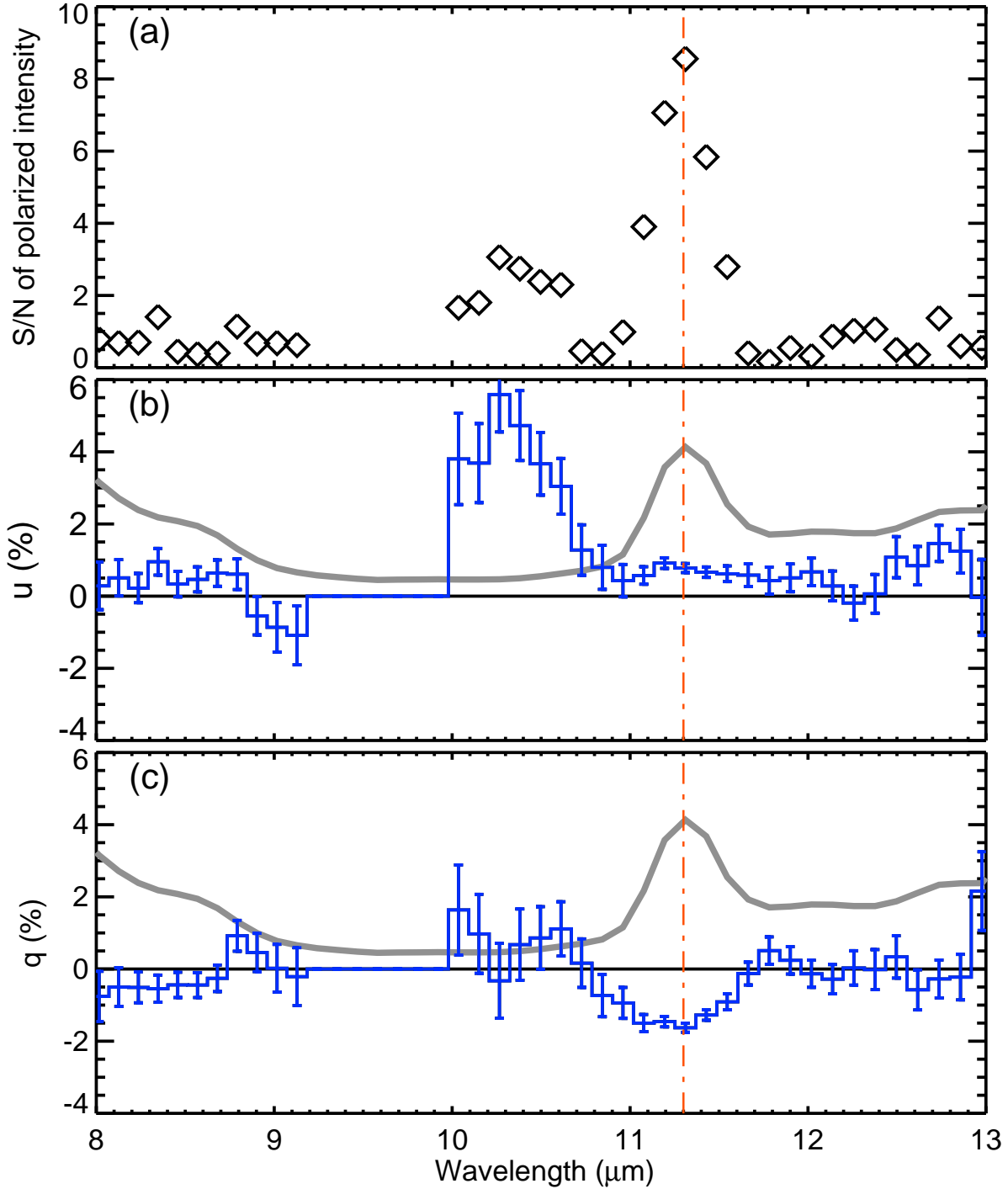}
\caption{(a) Signal-to-noise (S/N) ratio of polarized intensity of the NW nebula in Fig.\,\ref{fig:2}. The 11.3$\mum$ feature has around 9$\sigma$ detection. 
(b) Stokes $u (U/I)$. (c) Stokes $q (Q/I)$. 
The thick gray lines in (b) and (c) are the scaled intensity spectrum. 
The dashed line denotes the position 11.3$\mum$ PAH emission feature.     
The 10.3$\mum$ detected polarization feature mainly comes from in Stokes $u$, while the 11.3$\mum$ polarization mainly from Stokes $q$. The difference implies that they originate in different dust populations.}
\label{fig:3}
\end{figure}

We present in Fig.\,\ref{fig:2} mid-IR intensity and polarization spectra of the NW nebula covering the 8.0--13.0$\mum$ wavelength range.
The PAH emission features are clearly seen in Fig.\,\ref{fig:2}a, including the in-plane C--H bending feature at 8.6$\mum$ and
the out-of-plane C--H bending features
at 11.3 (solo-CH), 12.0 (duet-CH), and 12.7$\mum$ (trio-CH) \citep{1989ApJS...71..733A}.
The relative strengths of the PAH features 
depend on the size, structure, and charging
of PAHs \citep{1999A&A...352..659A, 2007ApJ...657..810D}, 
and the physical conditions of
the region where they are found (\citealt{1994ApJ...427..822B}; \citealt{2001ApJ...548..296W}).
By fitting the spectrum of MWC 1080 obtained with the {\it Infrared Space Observatory} (ISO), which exhibits a more complete set of PAH emission features but mixed emission from both stars and the nebula, \cite{2017ApJ...835..291S} determine that the best fit of PAHs in this environment are mostly neutral and large. 
Exposed to the energetic photons from
a B0e star with an effective temperature 
of $\sim$30,000$\K$ \citep{2008ApJ...673..315W},
small PAHs are probably unstable, and the only PAHs that survive are ones that are large and cata-condensed in structure.

The polarization spectrum shown together with 1$\sigma$ error bars of the NW nebula is presented in Fig.\,\ref{fig:2}b.
Two spectral regions show significant polarization. 
One is in the range of 10.9--11.7$\mum$, with p$_{11.3}$=1.9$\pm$0.2\% and position angle of 77.2\degr$\pm$3.2$\degr$, the latter indicated by the solid red lines superimposed on the intensity map of MWC 1080 in Fig.\,\ref{fig:1}. 
There is good consistency among the four separate measurements listed in Table \ref{tab:1}.
The other significant polarization is at 10.0--10.7$\mum$, peaking around 10.3$\mum$, and with p$_{10.3}$=5.4$\pm$1.6\% and position angle of 46.7\degr$\pm$8.2\degr. While the 10.3$\mum$ polarization feature is seemingly higher than that at 11.3$\mum$, the statistical significance of the results needs to be considered.
We note here that the detection of polarization near 10.3$\mum$ is only marginally significant, being barely 3$\sigma$, whereas the 11.3$\mum$ detection is robust, being around 9$\sigma$ (Fig.\,\ref{fig:3}a).
The 3$\sigma$ upper limits for the polarization at 8.6, 12.0 and 12.7$\mum$ are 1.9\%, 2.0\%, and 2.8\%, respectively (Fig.\,\ref{fig:2}b). 
While, theoretically, the emission at longer wavelengths is expected to have higher polarization (SD09), the 8.6, 12.0 and 12.7$\mum$ PAH emission features are much weaker than the 11.3$\mum$ feature and the lack of detection of the polarization in these regions is not surprising. 

Examining Stokes $u$ and $q$ explicitly (Fig.\,\ref{fig:3}) demonstrates that the 10.3$\mum$ polarization is apparent mainly in Stokes $u$, while the 11.3$\mum$ polarization is mainly in Stokes $q$. This difference implies a different origin for the 10.3$\mum$ if real and 11.3$\mum$ features, almost certainly indicating that they originate in different dust populations. 

\section{Discussion}\label{sec:dis}

The polarization feature between 10.9 and 11.7$\mum$ is well correlated in wavelength position with the 11.3$\mum$ PAH emission feature (Fig.\,\ref{fig:3}), and we conclude that the polarized emission is indeed due to PAH molecules. The peak value $p_{11.3}\simeq2\%$ is consistent with the previous PAH polarization search by \cite{1988A&A...196..252S} who established an upper limit of 3\% on its polarization in the Orion Bar.
The polarization position angle of the feature has an offset angle of $\sim$60\degr from the position angle of the projected illumination direction from the star to the nebula ($\sim$315\degr), as shown in Fig.1.

\subsection{Numerical Calculations using SD09}
We adopt the up-to-date SD09 models to model the degree of polarization in the environment of a reflection nebula. SD09 calculated the polarization for randomly oriented PAHs. They considered the intramolecular vibrational-rotation energy transfer (IVRET) process, which allows the efficient energy exchange between rotation and vibration.
SD09 model the polarization of PAH emission by introducing two parameters, $\gamma_{\rm 0}=T_{\rm rot}/T_{\rm 0}$ and $\gamma_{\rm ir}=T_{\rm rot}/T_{\rm ir}$, where $T_{\rm rot}$ is the rotational temperature determined by gas-grain interactions, photon absorption and IR emission, $T_{\rm 0}$ is the vibrational grain temperature that determines internal alignment prior to UV absorption, and $T_{\rm ir}$ is the temperature during IR emission. 
Alignment by an external magnetic field is ignored in their model, and the polarization degree of PAH emission features is essentially determined by two parameters, $\gamma_{\rm ir}$ and $\gamma_{\rm 0}$. Since the internal alignment temperature $T_{\rm ia}\equiv T_{\rm ir}$ is fixed for the different PAHs, and $T_{\rm 0}$ is determined by the radiation field, the polarization degree is determined by $T_{\rm rot}$. Consequently, the knowledge of $T_{\rm rot}$ is critical for achieving realistic predictions for the polarization of PAH emission. 

Collisions with gas neutrals and ions, UV absorption and subsequent IR emission, and electric dipole emission contribute to the rotation of PAHs (see \citealt{1998ApJ...508..157D}; \citealt{2010ApJ...715.1462H}). 
We assume in a reflection nebula that the gas temperature $T_{\rm gas}$=100 $\K$, gas density $n_{\rm H}=10^{3}$ ${\rm cm}^{-3}$, and the radiation field parameter $U=10^{3}$. We consider the PAH geometry as in \cite{1998ApJ...508..157D} and a typical grain size of $a$=7.5$\,\rm \AA$ (200 carbon atoms) in which most of the PAH mass is concentrated as in the ISM (SD09). We then carry out simulations of PAH dynamics using the Langevin code \citep{2010ApJ...715.1462H} and calculate the degree of polarization.
However, the values of polarization obtained using the SD09 model with the consideration of different hydrogen ionization fractions $x_{\rm H}$, i.e., $n_{\rm H^{+}}$/$n_{\rm H}$ ($n_{\rm H^{+}}$ is the number density of ionized hydrogen), are less than 0.5\% (Table \ref{tab:2}), much smaller than our measured value of $\sim$2\%.
It appears that other mechanisms that can enhance the PAH alignment need to be considered.     

\subsection{Alignment with Magnetic Fields}

SD09 discussed two possibilities that can significantly enhance the polarization of PAH emission, including suprathermal rotation (i.e., the PAH rotational temperature $T_{\rm rot}$ is much higher than the gas temperature $T_{\rm gas}$) and perfect internal alignment (i.e., the molecule principal axis is aligned with the angular momentum during both UV absorption and IR emission). 
The former is unlikely, since there are no obvious physical processes that can spin-up nanoparticles to suprathermal rotation rates (\citealt{1998ApJ...508..157D}; \citealt{2010ApJ...715.1462H}). 
On the other hand, perfect internal alignment can only be achieved at very low dust temperatures (T$_{\rm 0}$ of a few K) during the UV photon absorption, which requires very efficient energy exchange among the vibrational-rotational modes. 
This is also very unlikely due to quantum suppression that may occur in nanoparticles (\citealt{2016ApJ...831...59D}).  

Alternatively, we recognize that the enhanced polarization may arise from external alignment, i.e., the partial alignment of the angular momentum with an ambient magnetic field. 
Therefore, we have repeated the modeling incorporating the mechanism of resonance paramagnetic relaxation to align nanoparticles \citep{2000ApJ...536L..15L}. Subject to an external magnetic field, protons in aromatic hydrocarbons in the laboratory are known to experience stronger shielding (or deshielding) effects than regular hydrocarbons, because the $\pi$-electrons are delocalized and are free to circulate.
Astronomical magnetic fields can induce diamagnetic ring currents and polarizabilities in the $\pi$-electron clouds, resulting in coupling between the magnetic fields and two-dimensional PAHs, thereby forcing some alignment.
In our models, the magnetic field strength is assumed to be B=100 $\mu$G \citep{2012ARA&A..50...29C}. 
The calculation is described here briefly, with more details presented in Hoang (2017). 

We define $\hat{u}$-$\hat{v}$ to be the plane of the sky (see Hoang 2017).
We define $F_{\rm u,v}^{\|,\perp}$ to be the in-plane ($\|$) and out-of-plane ($\perp$) emission by a PAH molecule with the electric field $\bE$ in the $\hat{u}$-$\hat{v}$ plane. Then $I_{\rm u,v}^{\|,\perp}$ is the emission intensity from the PAH.
The flux $F_{\rm u,v}^{\|,\perp}$ depends on: 1) $f_{\rm LTE}(\theta,J)$, 
the probability distribution of the principal axis of the PAH plane being aligned with the grain angular momentum $\bJ$ (LTE stands for local thermal equilibrium); and 2) $f_{\rm J}$ ($\bJ$), the probability distribution of the angular momentum J being aligned with the direction of the magnetic field.  
The grain angular momentum $\bJ$ has spherical angles $\theta$ and $\phi$. $\beta$ is the nutation angle between $\bJ$ and the principal axis of the grain. The emission $I_{\rm w}$, with $w=(u,v)$, is obtained by integrating over the distribution functions 
 \begin{eqnarray}
I_{w}^{\|,\perp}(\alpha) &=& \int_{J} f_{J}(\bJ)d\bJ\int_{0}^{\pi}f_{LTE,0}(\theta_{0},J)d\theta_{0} \nonumber\\
&&\times \int_{0}^{\pi}f_{\rm LTE,ir}(\theta, J)d\theta
 A_{\star}(\beta,\theta_{0})F_{w}^{\|,\perp}(\beta, \phi,\theta,\alpha),\label{eq:Istar_uv}
\end{eqnarray}
where $A_{\star}$ is the cross-section of UV absorption as given in SD09. 
$f_{\rm LTE,0}$ and $f_{\rm LTE,ir}$ describe the thermal fluctuations of the principal axis of PAHs before UV absorption and during IR emission.
$f_{\rm LTE}(\theta,J)$ can be described by the Boltzmann distribution \citep{1997ApJ...484..230L} with $\int_{0}^{\pi}f_{\rm LTE}(\theta, J){\rm sin}\theta d\theta =1$. To simplify the calculation and derive the maximum value of polarization, we assume that the magnetic field is parallel to the stellar incident radiation direction.
We simulate the distribution of angular momentum from the Langevin equation assuming the ergodic system approximation and compute numerically the intensity of radiation using Equation (\ref{eq:Istar_uv}). 
The resulting degree of polarization with the viewing angle $\alpha$ for in-plane and out-of-plane modes is 
\begin{equation}
p^{\|,\perp}(\alpha)=\frac{I_{u}^{\|,\perp}(\alpha)-I_{v}^{\|,\perp}(\alpha)}{I_{u}^{\|,\perp}(\alpha)+I_{v}^{\|,\perp}(\alpha)},\label{eq:pol}
\end{equation}

We show the modeling results in Table \ref{tab:2} for the reflection nebula, which includes the ratio of the rotational temperature $T_{\rm rot}$ and the gas temperature $T_{\rm gas}$, the degree of alignment of the angular momentum with the magnetic field $Q_{\rm J}$, and the estimated polarization for the different hydrogen ionization fractions $x_{\rm H}$ with a viewing angle $\alpha$=90\degr.
The polarization varies from $0.14\%$ to $0.34\%$ in model a, increasing to $0.87\%$--$2.1\%$ when the external alignment is taken into account as in model b. 
In Table \ref{tab:2}, the degrees of 
polarization $p$ computed for both models 
increase with the increasing hydrogen 
ionization fraction $x_{\rm H}$. 
The results suggest that the polarization of 
PAH emission is dominated by PAHs in regions 
with a higher fraction of hydrogen in ionic form. 
In regions with a higher $x_{\rm H}$ 
(i.e., higher $n_{\rm H^{+}}$/$n_{\rm H}$), 
there will be more electrons available to 
neutralize the PAH ions created by photoionization \citep{2005pcim.book.....T}.
Indeed, as shown in \cite{2017ApJ...835..291S}, 
the aromatic hydrocarbon emission features
observed in MWC~1080 are best modeled in terms of
a mixture of PAHs with $\simali$80\% being neutral
and $\simali$20\% being ionized.  
As mentioned earlier, the delocalized $\pi$-electrons
in neutral PAHs may play a crucial role in coupling 
neutral PAHs with the magnetic field.

Partial alignment of PAHs with the magnetic field at a level of $Q_{\rm J}\simeq 0.08$--0.1 (the averaged degree of alignment of angular momentum $\bJ$ with \bB) is required to reproduce the observed $\sim$2\% polarization fraction at 11.3$\mum$.
This scenario also accounts for the observation that the polarization angle is offset from the illumination direction. 
When PAHs are aligned with the magnetic field, even though only partially, the polarization direction of the out-of-plane mode emission is expected to be along the magnetic field (Hoang 2017).

\subsection{Relationship between Polarization Angles and the Ambient Magnetic Field}
Both L88 and SD09 predict that the polarization associated with emission arising from the out-of-plane vibration mode should be along the illumination direction, which contrasts with our observed polarization angle having a $\sim$60$\degr$ offset. 
We explore one possible explanation, namely, PAH alignment by an external magnetic field (Table \ref{tab:2}). 
If magnetic alignment is important, we expect the polarization direction to match that of the ambient magnetic field lines.  
Indeed, we find that the optical polarimetry observations of 
 background stars within a few degrees on the sky from MWC1080 (serveral hundred parsecs in distance), indicates a fairly uniform optical polarization position angle of $\sim$80$\degr$, which suggests a smooth interstellar magnetic field threads the whole region (\citealt{2001AJ....122.3453M}; \citealt{2002A&A...387.1003M}; \citealt{2000AJ....119..923H}).
The value of the position angle agrees well with our measured polarization position angle of 77.2$\pm$3.2$\degr$ at 11.3$\mum$, which supports the hypothesis that PAHs are at least partially aligned with the ambient interstellar magnetic field threading the nebula and its neighborhood. 

Nevertheless, the emission and alignment of PAHs depend on local astrophysical conditions and the detailed properties of PAHs, especially their sizes. Nanoparticles with radii $\lesssim$10$\,\rm \AA$  are thought to be 
negligibly polarized with the greatest quantum suppression of alignment \citep{2016ApJ...831...59D}. 
Based on our results, it seems that other physical processes such as Faraday rotation braking that facilitate the alignment of nanoparticles need to be considered \citep{2016MNRAS.457.1626P}, since it is evident that the starlight anisotropy scheme alone in L88 is not sufficient to explain the measured high level of polarization. 

It is also worth noting that, given their abundances and small sizes, emission by rapidly spinning PAHs is widely believed to be the origin of AME in the 10--60 GHz frequency range \citep{1998ApJ...508..157D,1998ApJ...494L..19D}. 
If true, the considerable alignment of PAHs as suggested by our detection, naturally produces polarized spinning dust emission for which the polarization level is proportional to the degree of alignment of the PAH angular momentum with the magnetic field at $\sim$GHz frequencies \citep{2013ApJ...779..152H}. It implies that polarized emission from spinning Galactic-foreground PAHs can indeed constitute an obstacle to the detection of the CMB B-mode signal.

\subsection{Marginally Detected 10.3$\mum$ Polarization Feature} 
As shown in Fig.\,\ref{fig:3}a, we have a barely significant polarization detection at 10.3$\mum$ (3$\sigma$). 
If real, the different behaviors of the 10.3 and 11.3$\mum$ features in Stokes $u$ and $q$ (Fig.\,\ref{fig:3}) suggest that they originate in different dust populations. Therefore, it does not affect our interpretation of the high S/N (9$\sigma$) polarization detection at 11.3$\mum$, our main focus of this work. 
We do note, however, that there is no distinct Stokes I spectral fingerprint coinciding with the 10.3$\mum$ polarization. 
It is unlikely that the well-known silicate feature or one of its variants can account for this polarization, since the silicate absorptive polarization profiles are broad, spanning the entire 8--13$\mum$ region (e.g., \citealt{1993A&A...280..609H}; \citealt{2000MNRAS.312..327S}; \citealt{2017MNRAS.465.2983Z}) rather than relatively narrow and sharp as the feature we see here. 
Other possibilities, including nanoparticles with silicate \citep{RevModPhys.85.1021} or metallic Fe compositions, e.g., hygrogenated iron nanoparticles \citep{2017MNRAS.466L..14B}, might be worth investigating \citep{2016ApJ...821...91H} if further observations confirm and gain insight into the feature.

\section{Summary}

We report the unambiguous detection of polarized PAH emission at 11.3$\mum$ with a position angle of 77.2$\pm$3.2$\degr$ and polarization degree of 1.9$\pm$0.2\%, which confirms the PAH hypothesis that PAH molecules can indeed emit polarized light. 
The detection of polarization indicates that the alignment of PAHs is considerable. We find that the starlight anisotropy scheme alone is not sufficient to account for this polarization. The PAHs are at least partially aligned by the ambient magnetic field threading this young stellar region and its neighborhood, a conclusion strongly supported by the fact that the measured polarization angle is identical to the large-scale interstellar magnetic field spanning this region. This observation could have important consequences for the accurate estimate of Galactic foreground polarization, a consideration relevant to current goals to detect the CMB B-mode signal. 
We expect future polarimetry observations, e.g., with SOFIA/HAWC+ and GTC/CanariCam, covering the complete suite of PAH emission features (e.g., the 6.2$\mum$ band dominated by PAH cations and 3.3$\mum$ band by small PAHs) and various astrophysical environments, will deepen our understanding of the properties and alignment of PAHs, e.g., the effects of their sizes and charge states.

\acknowledgments
The authors are grateful for the anonymous referee's inspiring comments, which helped to improve the manuscript. The authors are grateful to the GTC staff for their outstanding support of the commissioning and science operations of CanariCam. We thank Ion Ghiviriga, Bruce Draine and Uma Gorti for their helpful discussions. 
C.M.T. acknowledges support from NSF awards AST-0903672, AST-0908624 and AST-1515331. A.L. is supported in part by NSF AST-1109039 and NNX13AE63G.
E.P. acknowledges the support from the AAS through the Ch\'retien International Research Grant and the FP7 COFUND program- CEA through an enhanced-Eurotalent grant, and the University of Florida for its hosting through a research scholarship. C.M.W. acknowledges financial support from Australian Research Council Future Fellowship FT100100495.

\clearpage

\begin{deluxetable}{cccccccc}
\tabletypesize{\scriptsize}
\tablecaption{Observing log \label{tab:1}}
\tablecolumns{8}
\tablewidth{0pt}
\tablehead{
\colhead{UT Date} &
\colhead{Target} &
\colhead{RA} & \colhead{DEC} &
\colhead{Integration } & \colhead{Airmass} &\colhead{$p_{11.3}$\tablenotemark{a}} & \colhead{$\theta_{11.3}$\tablenotemark{a}} \\
\colhead{} & \colhead{} &
\colhead{J2000} & \colhead{J2000} & \colhead{seconds} &\colhead{}& \colhead{\%} & \colhead{\degr}
}
\startdata
2015 Jul 31 & NW nebula & 23 17 25.18 & 60 50 44.56 & 993&1.18--1.23&2.46(0.42) &80.2(4.8)\\
2015 Aug 5 & NW nebula & 23 17 25.18 & 60  50 44.56 & 993&1.18--1.20&1.48(0.44) &88.7(8.2)\\
2015 Aug 5 & NW nebula & 23 17 25.18 & 60  50 44.56 & 993&1.21--1.28 &1.37(0.38) &72.0(7.7)\\
2015 Aug 7 & NW bebula & 23 17 25.18 & 60  50 44.56 & 993&1.20-1.27 &1.95(0.35) &70.2(5.1) \\
\enddata
\tablenotetext{a}{Values in parentheses are 1$\sigma$ uncertainties of measurements. All position angles are calibrated east from north.}

\end{deluxetable}

\begin{deluxetable}{ccccccc}
\tabletypesize{\scriptsize}
\tablecaption{Different models and polarization at 11.3$\mum$ \label{tab:2}} 
\tablecolumns{7}
\tablewidth{0pt}
\tablehead{
\colhead{ $x_{\rm H}$} &
\colhead{$T_{\rm rot}/T_{\rm gas}$} &
\colhead{$\gamma_{\rm ir}$} & \colhead{$\gamma_{\rm 0}$} &
\colhead{$p(\%)$\tablenotemark{a} } & \colhead{Q$_{\rm J}$} &\colhead{$p(\%)$\tablenotemark{b}} 
}
\startdata
0.001&0.62&1.54 & 0.31 & 0.14 & 0.058 & 0.87 \\
0.003 & 0.707 &1.76 &0.35 &0.19 & 0.069& 1.1 \\
0.005 & 0.76 &1.90 & 0.38 & 0.22 & 0.076 & 1.6 \\
0.010 & 0.94 &2.35 & 0.47 & 0.34 & 0.088 & 2.1 \\
\enddata
\tablenotetext{a}{Predicted polarization using the SD09 model assuming randomly oriented PAH angular momentum $\bJ$.}
\tablenotetext{b}{Predicted polarization with the external alignment of the grain angular momentum $\bJ$ with the magnetic field $\bB$ considered.}
\tablecomments{See Section 4.2 for the definitions of these parameters.}
\end{deluxetable}

\end{document}